# Single nanowire solar cells beyond the Shockley-Queisser limit

Peter Krogstrup[1,*], Henrik Ingerslev Jørgensen[2,*], Martin Heiss[3,*], Olivier Demichel[3], Jeppe V. Holm[2], Martin Aagesen[2], Jesper Nygard[1], Anna Fontcuberta i Morral[3,†]

[1] Center for Quantum Devices, Nano-Science Center, Niels Bohr Institute, University of Copenhagen, Denmark

[2] SunFlake A/S, Nano-Science Center, Universitetsparken 5, DK-2100 Copenhagen, Denmark

[3] Laboratoire des Matériaux Semiconducteurs, Ecole Polytechnique Fédérale de Lausanne, 1015 Lausanne, Switzerland

*equal contribution

**Light management is of great importance to photovoltaic cells, as it determines the fraction of incident light entering the device. An optimal pn-junction combined with an optimal light absorption can lead to a solar cell efficiency above the Shockley-Queisser limit. Here, we show how this is possible by studying photocurrent generation for a single core-shell p-i-n junction GaAs nanowire solar cell grown on a silicon substrate. At one sun illumination a short circuit current of 180 mA/cm$^2$ is obtained, which is more than one order of magnitude higher than what would be predicted from Lambert-Beer law. The enhanced light absorption is shown to be due to a light concentrating property of the standing nanowire as shown by photocurrent maps of the device. The results imply new limits for the maximum efficiency obtainable with III-V based nanowire solar cells under one sun illumination.**

Nanowire based solar cells hold promise for third generation photovoltaics and for powering nanoscale devices.[1,2] In the avenue of third generation photovoltaics, solar cells will become cheaper and more efficient than current devices, in particular a cost reduction may be achieved by the possibility of rationalizing the material use through the fabrication of nanowire arrays and radial p-n junctions.[3,4,5] The geometry of nanowire crystals is expected to favour elastic-strain relaxation, giving a major freedom for designing new compositional multi-junction solar cells[6] grown on mismatched materials.[7,8] Till today, the efficiencies of nanostructured solar cells have indeed been increasing with time up to 10%, owing to the improvement of material and device concepts.[9,10,11,12,13,14]

Light absorption in standing nanowires is a complex phenomenon, with a strong dependence on the nanowire dimensions and absorption coefficient of the raw material.[15,16,17,18] In low absorbing micro-wire arrays such as Si, light absorption is understood via ray optics or by the calculation of the integrated local density of optical states of the nanowire film.[19,20] Interestingly, when these arrays stand on a Lambertian back reflector, an asymptotic increase in light trapping for low filling factors is predicted.[19] This is advantageous for the improvement of the efficiency-to-cost ratio of solar cells and has led to the demonstration of micro-wire arrays exhibiting higher absorption than the equivalent thickness textured film[19,21,22] Quite different is the case of nanowires whose diameters are smaller than or comparable to the radiation wavelength. In this case, optical interference and guiding effects play a dominating role on the reflectivity and absorption spectra. For low absorbing materials (such as indirect bandgap, Si) wave-guiding effects plays a key role[23,24] while highly absorbing semiconductors (such as direct bandgap, GaAs) exhibit resonances that increase the total absorption several-fold. Nanowires lying on a substrate also exhibit such resonances, often described by Mie theory,[25,26] though the total absorption rate is significantly lower.[27,28] Even though optical absorption in nanowires pertaining to an array has shown to be far more complex than in thin films, nanowire vertical arrays seem to be the most reasonable device proposal nowadays. An elegant device consists in a single standing nanowire solar cell, contacted on top by a transparent electrode and at the bottom through the substrate. While the characterization of single nanowires lying on a substrate is quite common, until today, there are no studies on standing single nanowires.

In this work we present experimental measurements on single GaAs nanowire solar cells as grown on a silicon substrate where the p-part is contacted through a highly doped substrate and n-part with a transparent top-contact (see figure 1a-c and ref 26). We find that light absorption in single standing nanowires is more than one order of magnitude more efficient than what would be predicted from Lambert-Beer law. We show measurements on two devices. The first device (figure 1) exhibits a short circuit current density of 180 mA/cm$^2$, when normalized to the projected area. This leads to an apparent solar conversion efficiency of 40%. The second device shows a short circuit current of 173mA/cm$^2$ and an apparent efficiency of 28%. The reason for these very high efficiencies is the mismatch between the absorption cross-section and the physical bounds of the nanowires, hinting at a very large absorption cross section. This work represents a critical step towards the next generation of nanowire based solar cells.

Current-voltage characteristics of the devices were measured in the dark and under AM 1.5G illumination. Experimental data of device 1 are shown in Fig. 1d. In the dark, the device exhibits typical diode behaviour with an ideality factor of 2.6. Under illumination, the diode curve is shifted downwards as a consequence of the photo-generation and separation of electron-hole pairs, giving a short-circuit current of 256 pA. The diameter of the nanowire is 425 nm corresponding to an apparent photo-generated current density of 180 mA/cm². The open circuit voltage $V_{oc}$ and fill-factor FF are respectively 0.43 V and 0.52, which should be improved by optimizing the resistivity, thickness of the doped layers and surface passivation.[29] The generated power at the maximum power point is 57 pW, corresponding to 40 mW/cm². Dividing the generated power density by the incident power density, the solar cell yields an apparent efficiency of 40%. In order to understand the extreme photon collection boost in free standing single GaAs nanowires, we use finite difference time domain method to model a 2,5 µm long nanowire embedded in SU-8 as a function of its diameter and of the wavelength of the plane wave radiation propagating along the nanowire axis.[30,31,32] In figure 2a, the wavelength and diameter dependence of the absorption rate of such a nanowire is shown. Note that the absorption is zero for wavelengths larger than 900 nm where the absorption coefficient of GaAs goes down to zero. Two dominant branches for low and high diameters are observed, corresponding to resonances similar to Mie resonances observed in nanowires lying on a substrate[25]. Light absorption in the standing nanowire is enhanced by a factor between 10 and 70 with respect to the equivalent thin film. Another way to express this enhancement in absorption is through the concept of an absorption cross-section. The absorption cross-section is defined as $A_{abs} = a\eta$, where $a$ is the physical cross-section of the nanowire and $\eta$ is the absorption efficiency. It is largely accepted that the absorption cross-section in nanoscale materials is larger than their physical size. In systems such as quantum dots, the absorption cross-section can exceed the physical size by a factor of up to 8.[33] We have calculated the absorption cross-section of the nanowires as a function of the nanowire diameter and incident wavelength (Fig. 2b).

The absorption cross-section is in all cases larger than the physical cross-section of the nanowire. It is interesting to note that the absorption of photons from an area larger than the nanowire itself is equivalent to a build-in light concentration, $C$. Light concentration has an additional benefit that it increases the open circuit voltage with a term $kTlnC$ and thereby increasing the efficiency.[34,35,36] The largest absorption cross section in figure 2b is $1.13 \times 10^6$ nm$^2$ at a nanowire diameter of 380 nm ($a=9.38 \times 10^4$ nm$^2$), corresponding to an overall built-in light concentration of about 12. Measurements of the external quantum efficiency (EQE) normalized by the physical area for both lying and standing nanowire devices are shown in figure 3a (see supplement for more details). Lying nanowires exhibit EQE values up to 2 due to Mie resonances[37], while for standing nanowires values of up to about 14.5 are reached. This further confirms that the absorption cross-section is several times larger than the apparent cross section of the wire, especially at wavelengths close to the bandgap.

To further understand the absorption boost in our devices, we have spatially mapped the photocurrent generated by a vertical nanowire device for three different excitation wavelengths: 488, 676 and 800 nm. The results presented in Figures 3b-d are deconvoluted with the point spread function of the diffraction limited laser spot. As seen in the figure a photocurrent from an area much larger than the size of the laser spot appears for all three wavelengths. A fit to the data allows estimating an effective absorption cross-section diameter of 1.2 μm (488 nm), 1.0 μm (676 nm) and 1.3 μm (800 nm) respectively. Hence, the absorption boost in our device is due to an unexpected large absorption cross-section of the vertical nanowire geometry. This is equivalent to a built-in light concentration of about 8, which is in good agreement to what our theory predicts. In addition, we speculate that the top contact geometry further contributes to the resonant absorption effect, thereby increasing the absorption cross-section and the boost in photo-generated current.

Finally, we put our results in perspective by comparing them to top-notch existing technologies and with the design principles for increasing efficiency. In Table 1 we have listed the record values for leading technologies such as single junction crystalline silicon and GaAs, as well as triple junction devices. The highest efficiency is obtained by the triple-junction solar cell (34.1%), with a short-circuit current of 14.7 mA/cm$^2$. In this case, the short-circuit current is maintained relatively low, as it has to be matched between the three cells connected in series and what gives the high efficiency is the increase in $V_{oc}$. The highest short-circuit current is obtained in a crystalline silicon (c-Si) solar cell, where light management techniques resulted into the boost of photo-generated current of 42.7mA/cm$^2$. The record efficiency recorded by Alta Devices with GaAs was obtained with a relatively thin film, few microns in contrast to few hundred microns in standard cells. This brings us to the discussion of what determines high efficiency. A solar cell operates at a voltage that maximizes the generated power, dictated by the values of the short-circuit current, FF and $V_{oc}$. Design towards higher efficiencies points to strategies for increasing these values.[40,41] The first two parameters concern mainly the device 'engineering', while the *ultimate* $V_{oc}$ is dictated by the thermodynamics of the solar energy conversion into electrical work. Within the Shockley-Queisser model, $V_{oc}$ is limited by the following terms:[36]

$$V_{oc} = \frac{E_g}{q}\left(1 - \frac{T}{T_{sun}}\right) - \frac{kT}{q}\left(ln\frac{\Omega_{emit}}{\Omega_{sun}} + ln\frac{4n^2}{C} - ln\, QE\right)$$

where $T$ and $T_{sun}$ are the temperature of the cell and of the sun respectively, $\Omega_{emit}$ and $\Omega_{sun}$ correspond to the solid angle of emission and collection, $n$ the refractive index of the material and QE is the emission quantum efficiency. The first term is related to the Carnot efficiency, which reduces $V_{oc}$ by ~5%. The second term corresponds to the entropic losses occurring in the work generation. The first entropy loss is due to the non-reciprocity in the angle of light absorption and emission. Light resonant structures such as nanowires can reduce the contribution of this term.[36] The second entropy loss takes account of the concentration factor, given by the refractive index and any external concentration: $V_{oc}$ increases by implementing light trapping strategies, with the additional benefit of increasing absorption close to the bandgap.[42,43] The last entropic term refers to the non-radiative losses. It can be reduced to zero by increasing the QE to 1. The impressive result on GaAs cells by Alta Devices was obtained thanks to increasing QE in a GaAs thin film to 1. Our results provide a further path for higher efficiency solar cells. Even though the electrical characteristics shown are not ideal yet, we observe a light concentration effect plus the significant increase of absorption rate close to the bandgap, similar to what is proposed by ref 36. These two effects are so that nanowire structures can reduce entropy in the conversion of solar energy into electrical work, thereby providing a path for increasing efficiency of solar cells. It is also important to note that the unexpected increase in the absorption cross-section enables to further separate the nanowires from one another, resulting in major cost reductions of the final device. Our experiments indicate that a good inter-wire distance would be around 1.2 μm. A nanowire solar cell consisting of nanowires similar to the device showed in Fig 1 positioned in a hexagonal array with a pitch of 1 μm (optical cross-section with a diameter of ~1.2 μm) would have an optical filling-factor of 1 and it would only use an equivalent material volume of a 546 nm film and exhibit a conversion efficiency of 4.6%. Using devices with smaller area, see left branch from Fig. 1, one could further reduce the amount of material used up to a factor 72. By improving the electrical characteristics of the pn junction, higher efficiencies could be obtained. Just considering an effective light concentration of 15, an array of GaAs nanowires with ideal characteristics would exhibit an efficiency of 33.4% , hereby overcoming the Shockley-Queisser limit for the planar GaAs solar cells illuminated by 1.5AM radiation, according to the discussion presented above[34,44] Even higher efficiencies could be achieved if the device design could be tailored for higher light concentration and QE. Note that axial junctions, which have the same junction area as the projected area would obtain the full benefit from such concentrator effect, and it would be possible to directly compare performance of GaAs nanowire solar cells under 1 sun with planar GaAs cells under 10 suns. We demonstrate here that single nanowire devices generate several-fold higher power than their projected area convey when they are standing upright, which also minimizes their footprint. It should be pointed out that if one were to build a single nanowire solar cell, then a flat lying nanowire would exhibit ~15 times lower power density compared to the standing nanowire device due to the light concentration effect. This enhancement in the energy conversion at the nanoscale makes them useful as energy harvesters with minimum footprint feeding other devices on the same chip. This is already the case for nanowire based pn junctions with non-ideal characteristics like the one shown in this work. Last but not least, the improvement in the photon collection renders them in general ideal as photo-detectors.

In conclusion, we have observed a remarkable boost in absorption in single nanowire solar cells that is related to the vertical configuration of the nanowires and to a resonant increase in the absorption cross-section. These results open a new avenue for third generation solar cells, local energy harvesters on nanoscale devices and photon detectors.

**Methods**:

**Nanowire growth**. Nanowires were grown on an oxidized (111) Si with 100 nm apertures using a self-catalysed method. Ga nominal growth rate was 900nm/h, substrate temperature 630°C and V/III ratio 4[45,46]. P-doping of core was achieved by adding a flux of Be during axial growth[47]. Cores were annealed for 10 min at 630°C. The shell was obtained at 465°C, growth rate of 300 nm/h and with a V/III ratio of 50. N-type doping was obtained by adding Si to the growth.

**Device fabrication and characterization.** SU-8 was spun-on the substrate at 4000rpm for 45s and cured with 1min UV light and 3min on a hotplate at 185°C. Then an etch-back with 1-3min oxygen plasma etch was performed to free the nanowire tip. The top contact was defined by e-beam lithography followed by evaporation of indium-tin-oxide (more details can be found in supplementary information). The bottom contact was obtained by silver glueing to the back-side of the wafer. Current-voltage characteristics were measured in the dark and under 1.5G conditions with a standard solar simulator (LOT – Oriel 150W Xe lamp) with a 1" beam diameter and an AM1.5G filter. Photocurrent map of the devices were collected by scanning the contacted sample under the laser spot focused with a 63x and 0.75 N.A. lens.

**Finite Difference Time Domain method simulations.** The absorption of standing 2.5 µm GaAs nanowires of different diameters standing on silicon and surrounded by SU-8 was calculated by solving Maxwell equations in three dimensions for an incident plane wave radiation normal to the substrate. The wave equation is solved in time domain following refs 30 and 27.

**Acknowledgements:**

This research has been funded by the ERC starting grant UpCon, by SNF through projects nr 137648, 143908 and NCCR-QSIT. AFiM thanks STI for the 2011 end-of-year fund for MiBoots robots used in the scanning photo-current experiment. AFiM and MH thank Anna Dalmau-Mallorqui and Franz Michael Epple for experimental support. We thank Claus B. Sørensen and Morten H. Madsen for assistance on MBE growth. This work was supported by the Danish National Advanced Technology Foundation through project 022-2009-1, a University of Copenhagen Center of Excellence, and by the UNIK Synthetic Biology project.

**Author contribution:**

PK grew the nanowire pn junctions. HIJ did the I-V characterisation and fabricated the device with help from JVH and MA. MH and OD performed the FDTD calculations. MH realized the photocurrent mappings and the external quantum efficiency measurements. AFiM and PK conceived and designed the experiments. AFiM, JN and MA supervised the project. AFiM, HIJ, PK and MH made the figures and wrote the manuscript. All authors discussed the results and commented on the manuscript.

**Author information:**

Correspondence and requests for materials should be addressed to anna.fontcuberta-morral@epfl.ch or krogstrup@fys.ku.dk.

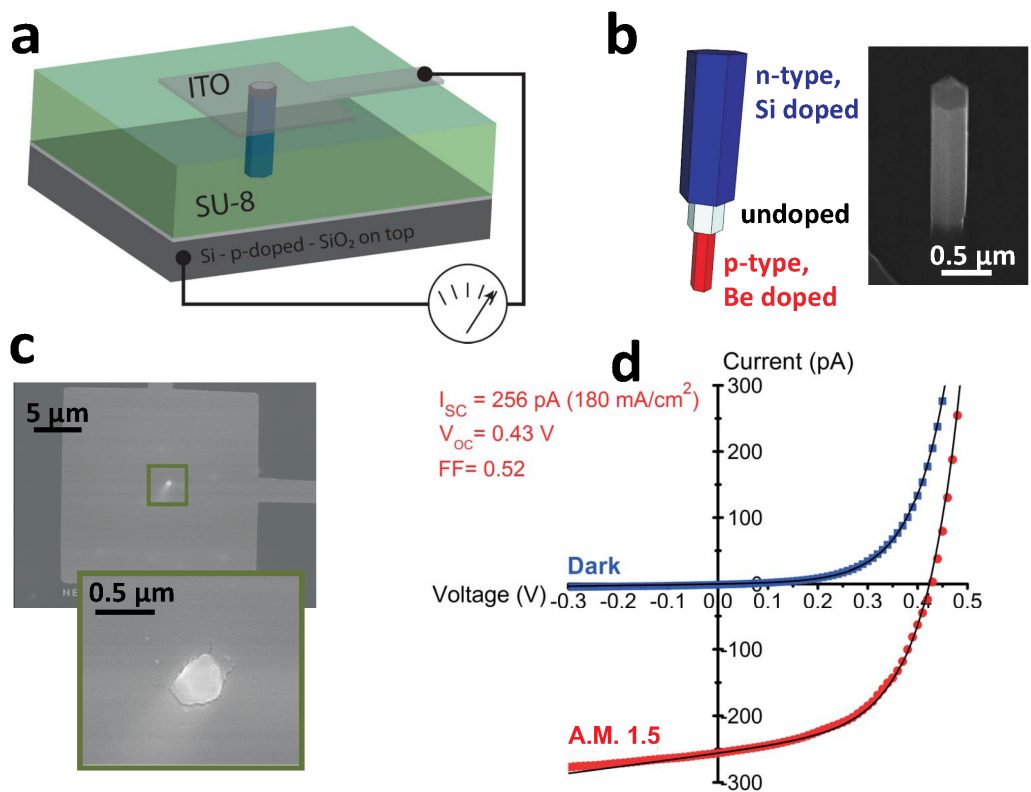

Figure 1. **Electrical characterization of a single nanowire solar cell (device 1). a**. Schematic of the vertical single nanowire radial p-i-n-device connected to a p-type doped Si wafer by epitaxial growth, **b** Left: Doping structure of the nanowire. The p-type doped core is in contact with the doped Si substrate and the n-type doped shell is in contact with the ITO is illustrated. Right: A typical SEM image of a nanowire from the same growth with a 30° angle from the vertical **c** Scanning electron micrographs of the device seen from the top electrode. The nanowire is ~2.5 μm high and has a diameter of about 425nm. **d** Current voltage characteristics of the device in the dark and under AM 1.5G illumination, showing the figure of merit characteristics.

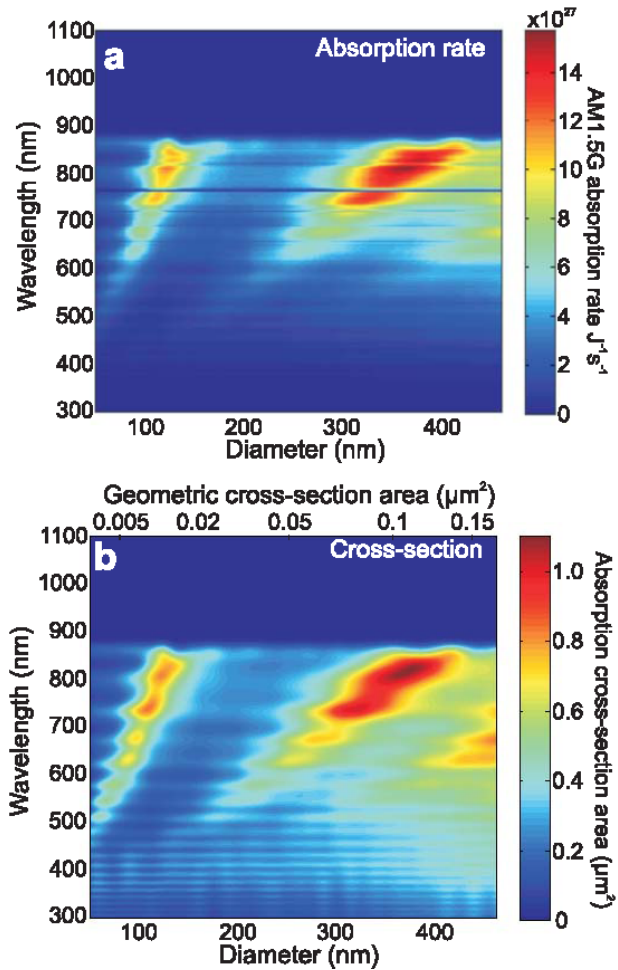

Figure 2. **Optical simulations of a single nanowire solar cell.** Simulations on the light absorption in a 2.5 μm standing GaAs nanowire that is fully embedded by SU-8 (n=1.67) on a Si substrate: the absorption rate of solar AM1.5G radiation in **a** and simulated absorption cross-section in **b** exhibit two main resonant branches, similar to Mie resonances observed in nanowires lying on a substrate. The periodic modulation with wavelength is a result of Fabry-Pérot interference in the polymer layer and not an artifact of the simulation.

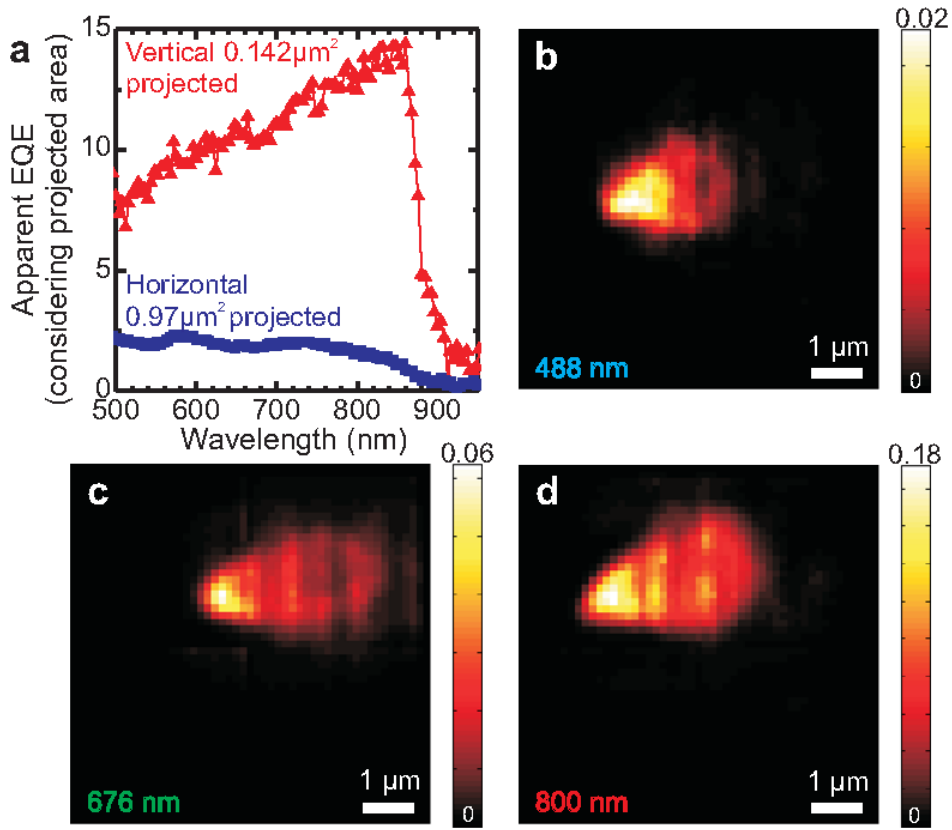

Figure 3. **. Optical characterization of a single nanowire solar cell (device 2). a**. External quantum efficiency (EQE) normalized by indicated projected area where vertical and horizontal nanowire solar cells are compared. As seen for the vertical standing solar cell a 15 fold increase in the photon collection is obtained close to the bandgap. The EQE becomes negligible for photon wavelengths below the bandgap of GaAs, meaning that there is no contribution from absorption in the Si substrate[47] (**b-d**) Scanning photocurrent measurements on our single vertical nanowire device for three different excitation laser wavelengths, normalized to the incident photon flux. The scale bar corresponds to 1 μm.

| Technology | $J_{SC}$ (mA/cm$^2$) | FF(%) | $V_{OC}$ (V) | Area (cm$^2$) | η (%) |
|---|---|---|---|---|---|
| c-Si | 42,7 | 82.8[38] | 0.706 | 4 | 25.0 |
| GaAs | 29.68 | 86.5 | 1.122 | 1 | 28.8[39] |
| Triple-junction | 14.57 | 86.0 | 3.014 | 30 | 37.7 |
| This work | 180 | 52 | 0.43 | 1.42x10$^{-9}$ | 40* |

**Table1.** Short-circuit current ($J_{SC}$), fill-factor, open-circuit voltage, area and efficiency of top-notch photovoltaic technologies compared with the standing nanowire configuration presented in this work. The low $V_{oc}$ and FF values indicate the potential for improvement of the nanowire cell presented, see text below for discussion (*apparent efficiency calculated with the projected area of the cell).